\documentclass[conference]{IEEEtran}

\usepackage{graphics,epsf,psfrag,amsmath,amssymb,amsfonts,verbatim,cite}
\usepackage[mathscr]{eucal}

\def\ACSP{/research/ACSP}

\newtheorem{theorem}{Theorem}

\newtheorem{lemma}{Lemma}

\newcommand{\Hmsc}{\mathscr{H}}
\newcommand{\Mmsc}{\mathscr{M}}

\newcommand{\Rmsc}{\mathscr{R}}

\newcommand{\Vmsc}{\mathscr{V}}
\newcommand{\Xmsc}{\mathscr{X}}

\newcommand{\Zmsc}{\mathscr{Z}}

\newcommand{\Rmbb}{\mathbb{R}}

\newcommand{\beq}{\begin{equation}}
\newcommand{\eeq}{\end{equation}}
\newcommand{\eps}{\epsilon}

\title{Variable-Rate Distributed Source Coding in the Presence of Byzantine Sensors}

\author{
\authorblockN{Oliver Kosut and Lang Tong}
\authorblockA{School of Electrical and Computer Engineering\\
Cornell University, Ithaca, NY 14853\\
Email: {\tt \{oek2,lt35\}@cornell.edu}}}

\begin{document}
\maketitle

\begin{abstract}
The distributed source coding problem is considered when the sensors, or encoders, are under Byzantine attack; that is, an unknown number of sensors have been reprogrammed by a malicious intruder to undermine the reconstruction at the fusion center. Three different forms of the problem are considered. The first is a variable-rate setup, in which the decoder adaptively chooses the rates at which the sensors transmit. An explicit characterization of the variable-rate minimum achievable sum rate is stated, given by the maximum entropy over the set of distributions indistinguishable from the true source distribution by the decoder.  In addition, two forms of the fixed-rate problem are considered, one with deterministic coding and one with randomized coding. The achievable rate regions are given for both these problems, with a larger region achievable using randomized coding, though both are suboptimal compared to variable-rate coding.

\vspace{1em} {\em Index Terms}---Distributed Source Coding. Byzantine Attack. Sensor Fusion. Network Security.
\end{abstract}

\section{Introduction}

Wireless sensor networks are vulnerable to various forms of attack. A malicious intruder could capture a sensor or a group of sensors and reprogram them, unbeknownst to the other sensors or the fusion center. The intruder could reprogram the sensors to work cooperatively to obstruct or defeat the goal of the network, launching a so-called Byzantine attack.

We refer to sensors that have been reprogrammed as \emph{traitors}, and the rest, which will behave according to the specified procedure, as \emph{honest}. Suppose there are $m$ sensors and at most $t$ traitors. Each time step, sensor $i$ is informed of the value of the random variable $X_i$. These random variables constitute a discrete memoryless multiple source with probability distribution $p(x_1\cdots x_m)$. Each sensor encodes its observation independently and transmits the codewords to a common decoder (the fusion center), which attempts to reconstruct the source values with small probability of error based on those transmissions. If there are no traitors, Slepian-Wolf coding \cite{Slepian&Wolf:73IT} can be used to achieve a sum rate as low as
\beq H(X_1\cdots X_m).\label{eq:swsumrate}\eeq
However, standard Slepian-Wolf coding has no mechanism for handling any deviations from the agreed-upon encoding functions by the sensors. Even a random fault by a single sensor could have devastating consequences for the accuracy of the source estimates produced at the decoder, to say nothing of a Byzantine attack on multiple sensors. 

Consider a two sensor example. If sensor 1 transmits at rate $H(X_1)$ and sensor 2 transmits at rate $H(X_2|X_1)$, their source sequences would normally be reconstructable using Slepian-Wolf. Since sensor 2 transmits at a rate below $H(X_2)$, the decoder must use the codeword from sensor 1 to decode $X_2$. Thus, if sensor 1 is a traitor, it can manipulate the decoder's estimate of $X_2$ to cause an error. Generalizing this, it will turn out that for most source distributions, the sum rate given in \eqref{eq:swsumrate} cannot be achieved if there is even a single traitor. We will present coding schemes that can handle Byzantine attacks, and give explicit characterizations of the achievable rates.

\subsection{Related Work}

The notion of Byzantine attack has its root in the Byzantine generals problem \cite{Lamport&Shostak&Pease:82ACM,Dolev:82} in which
a clique of traitorous generals conspire to prevent loyal generals from forming consensus. It was shown in \cite{Lamport&Shostak&Pease:82ACM} that consensus is possible if and only if less then a third of the generals are traitors.

Countering Byzantine attacks in communication networks has also been studied in the past by many authors.  See the earlier work of Perlman \cite{Perlman:88thesis} and also more recent review \cite{Zhou&Haas:99,Hu&Perrig:04}. An information theoretic network coding approach to Byzantine attack is presented in \cite{Ho&etal:04ISIT}. The problem of optimal Byzantine attack of sensor fusion for distributed detection is considered in \cite{Marano&Matta&Tong:06Asilomar}. Sensor fusion with Byzantine sensors was studied in \cite{Kosut&Tong:06Allerton}. In that paper, the sensors, having already agreed upon a message, communicate it to the fusion center over a discrete memoryless channel. Quite similar results were shown in \cite{Jaggi&etal:05ISIT}, in which a malicious intruder takes control of a set of links in the network. The authors show that two nodes can communicate at a nonzero rate as long as less than half of the links between them are Byzantine. This is different from the current paper in that the transmitter chooses its messages, instead of relaying information received from an outside source, but some of the same approaches from \cite{Jaggi&etal:05ISIT} are used in the current paper, particularly the use of randomization to fool traitors that have already transmitted.

\subsection{Fixed-Rate Versus Variable-Rate Coding}

In standard multiterminal source coding, each sensor is associated with a rate and an encoding function that transmits information at that rate. We will show that this fixed-rate setup is suboptimal for this problem, in the sense that we can achieve lower sum rates using a variable-rate scheme. By variable-rate we mean that the number of bits transmitted per source value by a particular sensor will not be fixed. Instead, each sensor has a number of different encoding functions, each with its own rate. The coding session is then made up of a number of transactions. In each transaction, the decoder decides which sensor will transmit information, and which encoding function it should use. Thus we require that the decoder have a reverse channel to transmit information back to the sensors, but it need only send the chosen encoding function index, which will be one of a fixed and small number. In other words, the reverse channel could have arbitrarily small capacity.

\subsection{Honest Sensor Error Requirement}

Classical Slepian-Wolf coding requires that the decoder produce perfect estimates of every source value. However, this is no longer possible under Byzantine attack. A traitor could choose to send gibberish to the decoder, in which case the decoder could never correctly decode the associated source values. However, a traitor could also act exactly like an honest sensor, in which case the decoder would never be able to identify it as a traitor. Thus, the decoder will not necessarily be able to produce an accurate estimate for every sensor, but neither will it be able to tell which of its estimates are inaccurate. As a compromise, the decoder will produce an estimate for every source value, but we only require that the estimates corresponding to the honest sensors are correct, even though the decoder may not know which those are. This requirement is reminiscent of that of \cite{Lamport&Shostak&Pease:82ACM}, in which the lieutenants need only perform the order given by the commander if the commander is not a traitor, even though the lieutenants might not know whether he is.

\subsection{Main Results}

The main results of this paper give explicit characterizations of the achievable rates for three different setups. The first, discussed in the most depth, is the variable-rate case, for which we give the minimum achievable sum rate. By definition, variable-rate coding involves varying the rates at which different sensors transmit. The choice of these rates will be based on ``run time'' events such as the source values and the actions of the traitors. Thus, there is no notion of an $m$-dimensional achievable rate region, since all we can say is that, no matter what happens, the total number of transmitted bits will not exceed a certain value. The second two setups are  fixed-rate, divided into deterministic coding and randomized coding, for which we do give $m$-dimensional achievable rate regions. We show that randomized coding yields a larger achievable rate region than deterministic coding, but we believe that in most cases randomized fixed-rate coding requires an unrealistic assumption. In addition, even randomized fixed-rate coding cannot achieve the same sum rates as variable-rate coding.

For variable-rate coding, the minimum achievable sum rate is given by
\beq\sup_{q\in Q}H_q(X_1\cdots X_m)\label{eq:introsumrate}\eeq
where $H_q$ is the entropy with respect to the distribution $q$ and $Q$ is a set of distributions which depends on $t$, the number of allowed traitors. The explicit definition of $Q$ is given later, but intuitively $Q$ is the set of distributions such that if we simulated any distribution $q\in Q$ and handed the resulting source sequences to the decoder as if they had come from the sensors, then it would not be able to correctly identify a single traitor. For example, the source distribution $p$ is always in $Q$, because if the decoder receives source sequences that appear to come from the true distribution, it will not be able to know which sensors are the traitors. In fact, if $t=0$, $Q$ is made up of only the source distribution $p$, so \eqref{eq:introsumrate} becomes \eqref{eq:swsumrate}. In other words, this result matches the classical Slepian-Wolf result.

On the other hand, if $t=m-1$, then the decoder knows only that the one honest sensor will report source values distributed according to its single variable marginal distribution, so a traitor will not be detected if it also reports source values distributed according to its marginal distribution. Hence $q\in Q$ if $q(x_i)=p(x_i)$ for all $i$. It is easy to see that \eqref{eq:introsumrate} becomes
\beq H(X_1)+\cdots +H(X_m).\label{eq:tm1}\eeq
In effect, the decoder must use an independent source code for each sensor.

The fixed-rate achievable regions are based on the Slepian-Wolf achievable region. For randomized coding, the achievable region is such that for every subset of $m-t$ sensors, the rates associated with those sensors fall into the Slepian-Wolf rate region on the corresponding $m-t$ random variables. Note that for $t=0$, this is identical to the Slepian-Wolf region. For $t=m-1$, this region is such that for all $i$, $R_i\ge H(X_i)$, which corresponds to the sum rate in \eqref{eq:tm1}. The deterministic region is similar, except that every subset of $m-2t$ rates is required to fall into the corresponding Slepian-Wolf region.

\subsection{Randomization}\label{subsection:random}

Randomization plays a key role in defeating Byzantine attacks. As we have discussed, allowing randomized encoding in the fixed-rate situation expands the achievable region. In addition, the variable-rate coding scheme that we propose relies heavily on randomization to achieve small probability of error. In both fixed and variable-rate coding, randomization is used as follows. Every time a sensor transmits, it randomly chooses from a group of essentially identical encoding functions. The index of the chosen function is transmitted to the decoder along with its output. Without this randomization, a traitor that transmits before an honest sensor $i$ would know exactly the messages that sensor $i$ will send. In particular, it would be able to find fake sequences for sensor $i$ that would produce those same messages. If the traitor tailors the messages it sends to the decoder to match one of those fake sequences, when sensor $i$ then transmits, it would appear to corroborate this fake sequence, causing an error. By randomizing the choice of encoding function, the set of sequences producing the same message is not fixed, so a traitor can no longer know with certainty that a particular fake source sequence will result in the same messages by sensor $i$ as the true one. This is not unlike Wyner's wiretap channel \cite{Wyner:75BSTJ}, in which information is kept from the wiretapper by introducing additional randomness.

In both variable-rate and randomized fixed-rate coding, we assume that the traitors know nothing about randomness produced at an honest sensor. Of course, after the randomness has been transmitted, the traitors should have access to that information, which is what we assume in the variable-rate case. However, for the fixed-rate setup, there is no notion of a transmission order, so it would be meaningless to say that the traitors only know about the randomness ``after'' it has been transmitted. The only choice is to assume that the traitors never find out anything about the randomness. This might be a realistic assumption if the traitors are not able to monitor transmissions to the decoder, but we believe that in most cases it is not. Hence deterministic fixed-rate coding is more realistic.

The rest of the paper is organized as follows. 
In Section~\ref{section:model}, we formally give the variable-rate model and present the main result of the paper, which we prove in Section~\ref{section:proof}.
In Section~\ref{section:outerbound}, we give the rate regions for the fixed-rate setups and illustrate that fixed-rate coding is suboptimal.
Finally, in Section~\ref{section:conclusion}, we offer some future avenues for research.

\section{Variable-Rate Model and Result}\label{section:model}

\subsection{Notation}

Let $X_i$ be the random variable revealed to sensor $i$, $\Xmsc_i$ the alphabet of that variable, and $x_i$ the corresponding realization. A sequence of random variables revealed to sensor $i$ over $n$ timeslots is denoted $X_i^n$, and a realization of it $x_i^n\in\Xmsc_i^n$. Let $\Mmsc\triangleq\{1,\ldots,m\}$. For a set $s\subset\Mmsc$, let $X_s$ be the set of random variables $\{X_i\}_{i\in s}$, and define $x_s$ and $\Xmsc_s$ similarly. By $s^c$ we mean $\Mmsc\backslash s$. Let $T_\eps^n(X_s)[q]$ be the strongly typical set with respect to the distribution $q$, or the source distribution $p$ if unspecified. Similarly, $H_q(X_s)$ is the entropy with respect to the distribution $q$, or $p$ if unspecified. All variations on $\eps$, such as $\eps',\eps'',\dot{\eps}$, are assumed to go to 0 as $\eps$ goes to 0 and may appear without definition. It is meant that either the definition is discernible from context or the existence will be shown.

\subsection{Communication Protocol}

The transmission protocol is composed of $L$ transactions. In each transaction, the decoder selects a sensor to receive information from and selects which of $K$ encoding functions it should use. The sensor then responds by executing that encoding function and transmitting its output back to the decoder. For each sensor $i\in\Mmsc$ and encoding function $j\in\{1,\ldots,K\}$, there is an associated rate $R_{i,j}$. On the $l$th transaction, let $i_l$ and $j_l$ be the sensor and encoding function chosen by the decoder, and let $h_l$ be the number of times $i_l$ has transmitted prior to the $l$th transaction. Note that $i_l,j_l,h_l$ are random variables, since they are chosen by the decoder based on messages it has received, which depend on the source values. The $j$th encoding function for the $i$th sensor is given by
\[f_{i,j}:\Xmsc_i^n\times\Zmsc\times\{1,\ldots,K\}^{h_l}\to\{1,\ldots,2^{nR_{i,j}}\}\]
where $\Zmsc$ represents randomness generated at the sensor. Let $I_l\in\{1,\ldots,2^{nR_{i,j}}\}$ be the message received by the encoder in the $l$th transaction. If $i_l$ is an honest sensor, then $I_l=f_{i_l,j_l}(X_{i_l}^n,\rho_{i_l},J_l)$, where $\rho_{i_l}\in\Zmsc$ is the randomness from sensor $i_l$ and $J_l\in\{1,\ldots,K\}^{h_l}$ is the history of encoding functions used by sensor $i_l$ so far. If $i_l$ is a traitor, however, it may choose $I_l$ based on all sources $X_1^n,\ldots,X_m^n$, all previous transmissions $I_1,\ldots,I_{l-1}$ and polling history $i_1,\ldots,i_{l-1}$ and $j_1,\ldots,j_{l-1}$. In particular, it does not have access to the randomness $\rho_i$ for any honest sensor $i$.

After the decoder receives $I_l$, if $l<L$ it uses $I_1,\ldots,I_l$ to choose the next sensor $i_{l+1}$ and its encoding function index $j_{l+1}$. After the $L$th transaction, it decodes according to the decoding function
\[g:\prod_{l=1}^L \{1,\ldots,2^{nR_{i_l,j_l}}\}\to \Xmsc_1^n\times\cdots\times\Xmsc_m^n.\]

\subsection{Variable-Rate Problem Statement and Main Result}

Let $\Hmsc\subset\Mmsc$ be the set of honest sensors. Define the probability of error
$P_e\triangleq\Pr(X_\Hmsc^n\ne \hat{X}_\Hmsc^n)$
where $(\hat{X}_1^n,\ldots,\hat{X}_m^n)=g(I_1,\ldots,I_L)$. This will in general depend on the actions of the traitors. Note again that the only source estimates that matter are those corresponding to the honest sensors.

We define a sum rate $R$ to be \emph{$\epsilon$-achievable} if for every $\delta>0$ and sufficiently large $n$ there exists a code such that, for any choice of actions by the traitors, $P_e\le\eps$ and 
\beq\sum_{l=1}^L R_{i_l,j_l}\le R+\delta.\label{eq:ratecondition}\eeq
Note that $R_{i_l,j_l}$ depend on the sensor transmissions, so they are random variables. By \eqref{eq:ratecondition} we mean that for any messages sent by the sensors, we never exceed a sum rate of $R+\delta$. A sum rate $R$ is \emph{achievable} if it is $\eps$-achievable for every $\eps>0$. Let $R^*$ be the minimum achievable sum rate. Certainly then all $R>R^*$ are also achievable.

Some definitions will allow us to state our main result. Let
\[\Vmsc\triangleq\{s\subset\Mmsc:|s|=m-t\}.\]
This is the collection of all possible sets of honest sensors. For any $V\subset\Vmsc$, define
\beq Q(V)\triangleq\{q(x_1\cdots x_m):\forall s\in V,\ q(x_s)=p(x_s)\}.\label{eq:qv}\eeq
Let $U(V)\triangleq \bigcup_{s\in V}s$. Finally, define
\[Q\triangleq\bigcup_{V\subset\Vmsc:U(V)=\Mmsc}Q(V).\]
That is, $Q$ is the set of distributions $q$ such that for each $i$, there is a marginal distribution of $q$ of $m-t$ variables including $X_i$ that matches the corresponding marginal distribution of $p$. Thus, those $m-t$ sensors behave as if they were the set of honest sensors, since their sources are distributed correctly. Since every $i$ falls into such a set, every sensor looks like it could be honest.
\begin{theorem}
The minimum achievable sum rate is
\beq R^*= \sup_{q\in Q}H_q(X_1\cdots X_m).\label{eq:thm}\eeq
\end{theorem}

It can be shown that for $t=1$ and arbitrary $m$, \eqref{eq:thm} becomes
\beq R^*= H(X_1\cdots X_m)+\max_{i,i'\in\Mmsc}I(X_i;X_{i'}|X_{\{i,i'\}^c}).\label{eq:1traitor}\eeq
Relative to the Slepian-Wolf result, we see that we always pay a conditional mutual information penalty for a single traitor. Similar expressions can be found for $t=2$, $t=m-2$, and $t=m-1$ (the last given by \eqref{eq:tm1}). However, analytic expressions do not in general exist for $3\le t\le m-3$.

\section{Proof of the Variable-Rate Theorem}\label{section:proof}

\subsection{Converse}

We first show the converse. Let $\tilde{q}$ be the distribution $q$ that maximizes the entropy in \eqref{eq:thm}. For some $s$ with $|s|=m-t$, we can write $\tilde{q}=p(x_s)\tilde{q}(x_{s^c}|x_{s})$. Thus if the $s^c$ sensors are the traitors, they can simulate the conditional distribution $\tilde{q}(x_{s^c}|x_{s})$, the outcome of which, when combined with the true values of $X_s$, will produce a set of $X_1\cdots X_m$ distributed according to $\tilde{q}$. Since $\tilde{q}\in Q$, if the traitors act honestly with these fabricated source values, the decoder will not be able to correctly identify a single traitor, so it has no choice but to perfectly decode every value. To do this, it must receive at least $nH_{\tilde{q}}(X_\Mmsc)$ bits, which means $R^*\ge H_{\tilde{q}}(X_\Mmsc)$.

\subsection{Achievability Preliminaries}

Now we prove achievability. To do so, we will need the following definitions. For some $V\subset\Vmsc$, let
\begin{equation*}S_\eps^n(X_\Mmsc)[V]\triangleq\{x_\Mmsc^n\in\Xmsc_\Mmsc^n:\forall s\in V, x_s^n\in T_\eps^n(X_s)\}\end{equation*}
where $T_\eps^n$ is the strongly typical set. For $s,s'\subset\Mmsc$ and $x_{s'}^n\in\Xmsc^n_{s'}$, we define the conditional version
\begin{multline*}S_\eps^n(X_s|x_{s'}^n)[V] \triangleq\{x_s^n\in\Xmsc_s^n:\exists x_{(s\cup s')^c}^n\in\Xmsc_{(s\cup s')^c}^n: \\(x_s^nx_{s'}^nx_{(s\cup s')^c}^n)\in S_\eps^n(X_\Mmsc)[V]\}.\end{multline*}
The following lemma shows that $S_\eps^n$ is contained in a union of typical sets.

\begin{lemma}\label{lemma:ssubset}
Fix $s,s'\subset\Mmsc$ and $x_{s'}^n\in\Xmsc^n_{s'}$. Then
\[S_\eps^n(X_s|x_{s'}^n)[V]\subset\bigcup_{q\in Q(V)}T_{\eps'}^n(X_s|x_{s'}^n)[q].\]
\end{lemma}

\subsection{Coding Scheme Procedure}

We propose a multiround coding scheme. Each round is made up of $m$ phases. In the $i$th phase, transactions are made entirely with sensor $i$. In addition, all transactions in the first round are based on the first $k$ source values, transactions in the second round on the second $k$ source values, and so on. Each transaction in the $i$th phase will be associated with a target set chosen by the decoder of the form 
\beq
T_R(\hat{x}_s^k)\triangleq\bigcup_{q:H_q(X_i|X_s)\le R}T_{\eps'}^k(X_i|\hat{x}_s^k)[q]\label{eq:trdef}\eeq
with $s\subset\Mmsc$ to be defined, and $\eps'$ is as defined in Lemma~\ref{lemma:ssubset}. It takes about $kR$ bits to encode any sequence in this set, so we can think of $T_R(\hat{x}_s^k)$ as the set of all the sequences that can be decoded if a sensor has only sent $kR$ bits so far in the current phase. The strategy will be to slowly increase $R$, expanding $T_R(\hat{x}_s^k)$ until it contains the relevant source sequence.

The decoder will attempt to determine whether the source sequence is contained in $T_R(\hat{x}_s^k)$, and if so to decode it. Sensor $i$ will randomly choose from a number of encoding functions $f_1,\ldots,f_C$. Each of these encoding functions will be created by means of a random binning procedure and the codebooks revealed to both the sensor and decoder. Sensor $i$ will transmit up to $k(R+\dot{\eps})$ bits containing the index of the randomly chosen encoding function and its output. If there is exactly one source sequence in the target set that matches every value received so far from sensor $i$ in this round, call it $\hat{x}_i^k$. If there is more than one such sequence, we declare an error. If there is no such sequence, we conclude that the source sequence is not contained in the target set, increase $R$ by $\eps$, and do another transaction. Note that when $R\ge\log |\Xmsc_i|$, every sequence will be in $T_R(\hat{x}_s^k)$, so we will definitely decode the sequence or declare an error.

The collection $V\subset\Vmsc$ will always contain only those sets that could be the set of honest sensors. We begin by setting $V=\Vmsc$, and pare it down after each round based on new information. Define \hbox{$s_i\triangleq\{1,\ldots ,i\}\cap U(V)$}. Phase $i$ of any round is made up of the following steps.
\begin{enumerate}
\item If $i\not\in U(V)$, ignore $i$ and go to the next phase.
\item Otherwise, let $R=\eps.$
\item Receive up to $k(R+\dot{\eps})$ bits from sensor $i$, with target set $T_R(\hat{x}_{s_{i-1}})$. If possible, decode the sequence to $\hat{x}_i^k$ and go to the next phase. If not, increase $R$ by $\eps$ and repeat.
\item After phase $m$, let $V'\in\Vmsc$ be the largest subset of $V$ such that $\hat{x}_{U(V)}\in S_\eps^n(X_{U(V)})[V']$. Use $V'$ as $V$ in the next round. If there is no such $V'$, declare an error.
\end{enumerate}

\subsection{Code Rate}

It can be shown that the probability of error can made arbitrarily small if $C$, the number of encoding functions from which each sensor chooses randomly during each transaction, is sufficiently large. We can then make $k$ large enough that transmitting the index of the chosen encoding function takes negligible rate compared to transmitting its output. Thus in each phase we need only transmit $R+\dot{\eps}$ bits per symbol. Let $q_{\hat{x}}$ be the type of $\hat{x}_{U(V)}^k$. The total number of bits sent per symbol for the entire round is therefore at most
\begin{align}\sum_{i=1}^m &\inf_{q:\hat{x}_i^k\in T_{\eps'}^n(X_i|\hat{x}_{s_{i-1}}^k)[q]}H_q(X_i|X_{s_{i-1}})+\eps+\dot{\eps}\nonumber\\
&\le \inf_{q:\hat{x}_{U(V)}^k\in T_{\eps'}^n(X_{U(V)})[q]}\sum_{i=1}^mH_q(X_i|X_{s_i})+m(\eps+\dot{\eps})\label{eq:rate1}\\
&\le H_{q_{\hat{x}}}\left(X_{U(V)}\right)+m(\eps+\dot{\eps})\label{eq:rate2}\\
&\le \sup_{q\in Q(V')}H_q\left(X_{U(V)}\right)+\ddot{\eps} \label{eq:rate3}\\
&\le \sup_{q\in Q}H_q(X_\Mmsc)+\log |\Xmsc_{U(V)\backslash U(V')}|+\ddot{\eps}\label{eq:rate4}
\end{align}
where \eqref{eq:rate1} holds because the set of distributions $q$ such that $\hat{x}_{s_i}^k\in T_{\eps'}^n(X_{s_i})[q]$ contains the set of distributions $q$ such that $\hat{x}_{U(V)}^k\in T_{\eps'}^n(X_{U(V)})[q]$, and \eqref{eq:rate2} holds because $\hat{x}_{U(V)}$ is typical with respect to its own type. Because $\hat{x}_{U(V)}\in S_\eps^n(X_{U(V)})[V']$, by Lemma~\ref{lemma:ssubset}, for some $q\in Q(V')$, $\hat{x}_{U(V)}\in T_{\eps'}^n(X_{U(V)})[q]$. For this $q$, for all $x_{U(V)}\in\Xmsc_{U(V)}$,
$\left|q_{\hat{x}}(x_{U(V)})-q(x_{U(V)})\right|\le\frac{\eps'}{|\Xmsc_{U(V)}|}.$
Since the distributions are arbitrarily close, the entropies with respect to these distributions will be arbitrarily close, so \eqref{eq:rate3} holds.

If $U(V')=U(V)$, then the second term in \eqref{eq:rate4} is 0, so we can bound \eqref{eq:rate4} by $\sup_{q\in Q}H_q(X_\Mmsc)+\ddot{\eps}$. However, if $U(V)\backslash U(V')\ne\emptyset$, we cannot. Even so, since at least one sensor is eliminated whenever $U(V)\backslash U(V')\ne\emptyset$, this can only happen for at most $t$ rounds, after which we will have eliminated every traitor. Thus with enough rounds, we can always bound the sum rate by $\sup_{q\in Q}H_q(X_\Mmsc)+\ddot{\eps}$.

\section{Fixed-Rate Results}\label{section:outerbound}

Consider an $m$-tuple of rates $(R_1,\ldots,R_m)$, encoding functions
$f_i:\Xmsc_i^n\to\{1,\ldots,2^{nR_i}\}$
for $i\in\Mmsc$, and decoding function
\[g:\prod_{i=1}^m\{1,\ldots,2^{nR_i}\}\to\Xmsc_1^n\times\cdots\times\Xmsc_m^n.\]
Let $I_i\in\{1,\ldots,2^{nR_i}\}$ be the message transmitted by sensor $i$. If sensor $i$ is honest, $I_i=f_i(X_i^n)$. If it is a traitor, it may choose $I_i$ arbitrarily, based on all the sources $X_\Mmsc^n$. Define the probability of error $P_e\triangleq\Pr\big(X_\Hmsc^n\ne \hat{X}_\Hmsc^n\big)$
where $(\hat{X}_1^n,\ldots,\hat{X}_m^n)=g(I_1,\ldots,I_L)$. 

We say an $m$-tuple $(R_1,\ldots,R_m)$ is \emph{deterministic-fixed-rate achievable} if for any $\eps>0$ and sufficiently large $n$, there exist coding functions $f_i$ and $g$ such that, for any choice of actions by the traitors, $P_e\le\eps$. Let $\Rmsc_{\text{dfr}}\subset\Rmbb^m$ be the set of deterministic-fixed-rate achievable $m$-tuples.

Define an $m$-tuple to be \emph{randomized-fixed-rate achievable} in the same way as above, except we allow the encoding functions $f_i$ to be randomized. Let $\Rmsc_{\text{rfr}}\subset\Rmbb^m$ be the set of randomized-fixed-rate achievable rate vectors.

For any $s\subset\Mmsc$, let $\text{SW}(X_s)$ be the Slepian-Wolf rate region for the random variables $X_s$. For any integer $k\le m$, define
\[
\Rmsc_k\triangleq\{(R_1\cdots R_m):\forall s\subset\Mmsc,|s|=k:\\
(R_i)_{i\in s}\in \text{SW}(X_s)\}.\]
The following theorem gives the rate regions explicitly.
\begin{theorem}\label{thm:fixedrate}
The fixed-rate achievable regions are given by
\[\Rmsc_{\text{dfr}}=\Rmsc_{\max\{1,m-2t\}}\qquad\text{and}\qquad \Rmsc_{\text{rfr}}=\Rmsc_{m-t}.\]
\end{theorem}

We omit the proof of this, but we briefly illustrate that circumstances exist for which fixed-rate coding is suboptimal compared to variable-rate coding. Suppose $m=3$ and $t=1$. Recall from \eqref{eq:1traitor} that the variable-rate minimum achievable sum rate is given by
\begin{multline}R^{*}= H(X_1X_2X_3)+\max\{I(X_1;X_2|X_3),\\I(X_1;X_3|X_2),I(X_2;X_3|X_1)\}\label{eq:3rate}.\end{multline}
Suppose that $I(X_1;X_2|X_3)$ achieves this maximum. If the rate triple $(R_1,R_2,R_3)$ is randomized fixed-rate achievable, then $(R_1,R_2,R_3)\in\Rmsc_2$, which means $R_i+R_j\ge H(X_iX_j)$ for all $i,j\in\{1,2,3\}$. Thus
\begin{gather}
R_1+R_2+R_3\ge \frac{1}{2}\big[H(X_1X_2)+H(X_1X_3)+H(X_2X_3)\big]\nonumber\\
=H(X_1X_2X_3)+\frac{1}{2}\big[I(X_1;X_2|X_3)+I(X_1X_2;X_3)\big].
\label{eq:fixed3rate}
\end{gather}
If $I(X_1X_2;X_3)>I(X_1;X_2|X_3)$, \eqref{eq:fixed3rate} is larger than \eqref{eq:3rate}. Hence, for some source distributions, a larger sum rate is required for fixed-rate coding than variable-rate coding.

\section{Future Work}\label{section:conclusion}

Much more work could be done in the area of Byzantine network source coding. In this paper, we assumed that the traitors have access to all the source values, an assumption that was vital in our converse proofs. This is a significant assumption that may not be all that realistic. It would be worthwhile, though perhaps more difficult, to characterize the achievable rate region without this assumption, assuming that the traitors have access only to their own source values, or possibly degraded versions of those of the honest sensors.

Finally, we could consider Byzantine attacks on other sorts of multi-terminal source coding problems, such as the rate distortion problem \cite{Tung:thesis,Berger:Chapter78} or the CEO problem \cite{Berger&etal:96IT}.

\bibliographystyle{ieeetr}
{
\bibliographystyle{ieeetr}
\bibliography{\ACSP/Reference/Bibs/Journal,\ACSP/Reference/Bibs/Conf,%
\ACSP/Reference/Bibs/Book,\ACSP/Reference/Bibs/Misc,\ACSP/Reference/Bibs/ACSP-J,\ACSP/Reference/Bibs/ACSP-C}
}

\end{document}